\documentclass[aps, prd, twocolumn, amsmath, floats,floatfix, superscriptaddress, nofootinbib]{revtex4}
\usepackage{graphicx}
\usepackage{bm}
\usepackage{diagbox}
\usepackage{hyperref}
\usepackage[utf8]{inputenc}
\usepackage{subfigure,booktabs}
\usepackage{pifont}
\makeatletter\def\Hy@Warning#1{}\makeatother
\hypersetup{
  colorlinks=true,        
  linkcolor=blue,         
  citecolor=cyan,         
  urlcolor=cyan            
}
\usepackage{amsmath,amsfonts,amssymb}
\usepackage{times}
\usepackage[T1]{fontenc}
\usepackage{float}
\usepackage[normalem]{ulem}
\usepackage{multirow}
\usepackage{soul}

\newcommand{\be}{\begin{equation}}
\newcommand{\ee}{\end{equation}}
\newcommand{\bea}{\begin{eqnarray}}
\newcommand{\eea}{\end{eqnarray}}

\makeatletter
\newcommand*{\rom}[1]{\expandafter\@slowromancap\romannumeral #1@}
\makeatother

\makeatletter
\newcommand{\thickhline}{%
    \noalign {\ifnum 0=`}\fi \hrule height 1pt
    \futurelet \reserved@a \@xhline
}
\newcolumntype{"}{@{\hskip\tabcolsep\vrule width 1pt\hskip\tabcolsep}}
\makeatother
\begin{document}
\title{
NS-UNO: Neutron Star EoS Inference from an Unconstrained Number of Observations 
}

\author{Valéria Carvalho}
\email{val.mar.dinis@uc.pt}
\affiliation{CFisUC, 
	Department of Physics, University of Coimbra, P-3004 - 516  Coimbra, Portugal}
\affiliation{Nicolaus Copernicus Astronomical Center, Polish Academy of Sciences, Bartycka 18, 00-716, Warsaw, Poland}
\author{Márcio Ferreira}
\email{marcio.ferreira@uc.pt}
\affiliation{CFisUC, 
	Department of Physics, University of Coimbra, P-3004 - 516  Coimbra, Portugal}

\author{Micha{\l} Bejger}
\email{bejger@camk.edu.pl}
\affiliation{Nicolaus Copernicus Astronomical Center, Polish Academy of Sciences, Bartycka 18, 00-716, Warsaw, Poland}
\affiliation{INFN Sezione di Ferrara, Via Saragat 1, 44122 Ferrara, Italy}

\author{Constança Providência}
\email{cp@uc.pt}
\affiliation{CFisUC, 
	Department of Physics, University of Coimbra, P-3004 - 516  Coimbra, Portugal}
\date{\today}

\begin{abstract}
Future multimessenger observations of neutron stars (NS) are expected to substantially increase both the number and precision of astrophysical constraints on the equation of state (EoS) of dense matter. This motivates inference frameworks capable of accommodating a variable, non fixed number of observations while preserving the posterior information associated with each measurement.
In this work, we introduce NS-UNO, a Neural Posterior Estimation framework for NS EoS inference designed to accommodate an Unconstrained Number of Observations (UNO). NS-UNO combines a hierarchical DeepSets model with a conditional normalising flow, enabling a single trained model to perform inference from mass-radius observation sets of varying size, with each observation represented by a set of posterior samples.
We demonstrate accurate and well calibrated posterior reconstructions using a model trained jointly on piecewise polytropic and non-parametric Gaussian process EoS ensembles. The reconstruction improves as observations probe a broader range of NS masses, while remaining robust to variations in the number and precision of the observations. The model also generalises to EoSs outside the families used during training. Finally, we qualitatively demonstrate the framework on current multimessenger constraints from NICER and GW170817. NS-UNO provides a flexible and scalable approach to NS EoS inference, naturally suited to the increasingly diverse observational datasets expected from next generation multimessenger astronomy.
\end{abstract}

\maketitle
\section{Introduction}

Understanding the equation of state (EoS) of neutron star (NS) matter remains one of the central challenges in nuclear physics and astrophysics, as it governs the structure and observable properties of NS, including their masses, radii, and tidal deformabilities \cite{Haensel2007NeutronStars1,glendenning2012compact,AnderssonN2019}. Since matter at these densities cannot be reproduced in terrestrial laboratories, the EoS must be constrained indirectly by combining theoretical models of dense matter with astrophysical observations \cite{chatziioannou2024neutron}.

Over the past decade, multimessenger observations have substantially improved these constraints. Gravitational-wave (GW) detections of binary NS mergers, notably the GW170817 event \cite{Abbott:2018wiz} 
(for the current state of the LIGO-Virgo-KAGRA (LVK) Collaboration \cite{ALIGO2015,AdV2015,KAGRA2013} detections, see \cite{2026arXiv260527223T}) 
have constrained the tidal deformability of NSs \cite{annala2018gravitational}, while X-ray pulse-profile modelling with NS Interior Composition Explorer (NICER) has provided increasingly precise $M(R)$ measurements for several pulsars \cite{Riley_2019,Miller19,Fonseca:2021wxt,choudhury2024nicer,salmi2024nicer,mauviard2025nicer}. Together with theoretical constraints from chiral effective field theory ($\chi$EFT) at low densities \cite{tews2013neutron} and perturbative QCD (pQCD) at asymptotically high densities \cite{kurkela2010cold,komoltsev2022perturbative}, these observations have considerably narrowed the range of viable EoSs. Nevertheless, the behaviour of matter above approximately twice nuclear saturation density remains poorly constrained, leaving open questions regarding the possible existence of phase transitions and other exotic degrees of freedom.

Bayesian inference of the NS EoS has traditionally relied on Markov Chain Monte Carlo (MCMC) or nested sampling methods \cite{Sharma:2017wfu,Skilling:2006gxv}. While highly successful, these approaches require the posterior to be recomputed whenever new observations are incorporated or different EoS parametrizations are considered.
As future observatories such as the Einstein Telescope \cite{abac2026science} and Cosmic Explorer \cite{reitze2019cosmic} dramatically increase both the number and diversity of multimessenger observations, this repeated inference will become increasingly computationally expensive. More importantly, future inference frameworks must naturally accommodate observational datasets of variable size.

Simulation based inference (SBI) offers an attractive alternative by learning the posterior directly from simulated data. In particular, Neural Posterior Estimation (NPE) approximates the posterior distribution of the model parameters conditioned on the observations, enabling fast amortized posterior inference once the model has been trained. NPE has already demonstrated excellent performance in GW parameter estimation \cite{dax2021real,Dax:2024mcn} and has recently been applied to NS EoS inference \cite{carvalho2025neural,thakur2026amortized,brandes2024neural}.

In our previous work \cite{carvalho2025neural}, we introduced an NPE framework for NS EoS inference from synthetic mass, radius, and tidal deformability observations. Although it produced accurate and well-calibrated posterior reconstructions, the architecture assumed a fixed number of observations, each represented by a single point estimate. Consequently, it could not naturally process future observational datasets containing variable numbers of sources represented by posterior samples.

To the best of our knowledge, existing approaches are designed for fixed-size conditioning vectors rather than hierarchical observational sets in which both the number of observations and the posterior samples associated with each observation may vary, \cite{thakur2026amortized,bezerra2026neutron,zhao2026semi,Fujimoto:2024cyv}.

Therefore we introduce \textbf{NS-UNO} (\textit{Neutron Star EoS inference from an Unconstrained Number of Observations}), an NPE framework capable of handling both an arbitrary number of NS observations and the posterior samples associated with each observation. NS-UNO combines a hierarchical DeepSets model \cite{zaheer2018deepsets} with a Conditional Normalizing Flow (CNF) \cite{CNF2019arXiv191200042W}, mapping hierarchical observational sets of arbitrary size into a fixed-dimensional, permutation-invariant representation that conditions the posterior over EoSs. To reduce dependence on a specific EoS parametrization, the framework is trained and validated using both piecewise-polytropic \cite{ferreira2025conditional} and non-parametric Gaussian-process EoS ensembles \cite{annala2023strongly}.

We demonstrate that NS-UNO provides accurate and well calibrated posterior reconstructions while remaining robust to varying observational configurations. We investigate how reconstruction depends on the number of observations, NS mass distribution coverage, and uncertainty of the available observations, evaluate its generalization to previously unseen EoS families, and finally present a qualitative application to current NICER and the LVK GW170817 observations.

The remainder of this paper is organized as follows. Section~\ref{dataset} describes the datasets and the generation of the mock observations. Section~\ref{model} presents the NS-UNO architecture and its implementation. Section~\ref{results} discusses the reconstruction performance and generalization tests, while Sec.~\ref{conclusion} summarizes the main conclusions and outlines future directions.

\section{Dataset}
\label{dataset}

The objective of this work is to develop a model capable of inferring the NS EOS independently of any specific parametrization. To achieve this, we combine the two datasets introduced in our previous work \cite{carvalho2025neural}, rather than training separate models for each. This merged dataset exposes the network to a wide range of physically plausible EoSs, allowing us to assess its ability to generalize across fundamentally different descriptions of dense matter.

\textbf{Piecewise-polytropic (PT) EoSs}. This family is generated following the prescription of \cite{ferreira2025conditional}. Each EoS is described by five connected polytropic segments, providing a flexible parametrization capable of representing a broad range of stiffnesses and phase transition like behaviour. All models satisfy causality and support NSs with masses above $2\,M_\odot$. For each EoS, the Tolman Oppenheimer Volkoff (TOV) equations \cite{1939PhRv...55..364T,1939PhRv...55..374O} are solved to obtain the corresponding $M(R)$ relation, from which synthetic observations are generated by sampling NS masses and adding Gaussian observational uncertainties.

\textbf{Gaussian Process (GP) EoSs}. This family is based on the non-parametric framework developed in \cite{annala2023strongly,komoltsev_2023_10101447}. GP regression is used to interpolate between $\chi$EFT at low densities and pQCD at high densities, producing EoSs that satisfy causality and thermodynamic stability. The prior is further constrained using current astrophysical observations, including measurements of massive pulsars \cite{antoniadis2013massive,cromartie2020relativistic}, tidal deformability from GW170817 \cite{Abbott:2018wiz}, and $M(R)$ measurements from X-ray observations such as NICER \cite{shaw2018radius,steiner2018constraining,Fonseca:2021wxt}. Unlike the PT family, the GP ensemble does not assume an explicit functional form for the EoS, providing a complementary and highly model independent description of dense matter.

By combining PT and GP EoS families into a single training set, the framework learns from both parametric and non-parametric EoSs, spanning a broad range of physical behaviours and reducing dependence on any particular EoS representation.

\subsection{Generation of Mock Datasets}

\subsubsection{EoS representation}
After merging the PT and GP families, the final dataset contains 118\,419 EoSs. We use 90\% of the dataset for training and validation and the remaining 10\% for testing. The training and validation subset is further divided into 90\% for training and 10\% for validation, resulting in 106\,567 EoSs for training and validation and 11\,852 for testing.


Prior to training, the GP ensemble is filtered by requiring each EoS to support a NS with a maximum mass above $2\,M_\odot$. No likelihood-based filtering is applied, ensuring that the training set spans a broad range of physically viable EoSs.

Each EoS is represented by the pressure evaluated at twenty equally spaced baryon densities,
\begin{equation}
\mathbf{p} = [p(n_1),p(n_2),\ldots,p(n_{20})]\,, 
\end{equation}
where $n_1=0.13~\mathrm{fm}^{-3}$, $n_{20}=1.28~\mathrm{fm}^{-3}$ and $\Delta n=0.0605~\mathrm{fm}^{-3}$. The pressure values are transformed to $\log_{10}(p)$ before training to reduce the dynamic range and improve numerical stability.

\subsubsection{Mock observations} \label{sec:Mock_observations}

Unlike previous approaches, the proposed framework is designed to infer the EoS from an arbitrary number of NS observations. Consequently, the training data should not be tied to the characteristics of the current observational sample alone, but should also encompass plausible future scenarios.

The construction of the mock observations is therefore guided by four unknown aspects of present and future datasets:
\begin{enumerate}
    \setlength{\itemsep}{-0.25em}
    \item number of observed NSs,
    \item number of synthetic samples available for each observation,
    \item distribution of NS masses,
    \item observational uncertainties.
\end{enumerate}

To account for these unknowns, we randomly vary each of these quantities during training, allowing the model to learn a posterior that is robust to different observational configurations. The following subsections describe how each aspect is modelled.

For each EoS, we first generate a set of $N_{obs}$ synthetic NS observations. Stellar masses are sampled using a stratified scheme spanning the accessible mass range $1 M_\odot < M<M_{max}$(EoS) such that different mass regimes of the corresponding TOV sequence are represented during training. This sampling strategy is used only to ensure that different mass regimes are represented during training and is not intended to model the current astrophysical NS mass distribution.

The corresponding radii are obtained by interpolating the TOV $M(R)$ relation of the underlying EoS. Each synthetic observation is then represented by a cloud of correlated samples drawn from a bivariate Gaussian distribution,
\begin{equation}
\begin{pmatrix}
M_{o,s}\\
R_{o,s}
\end{pmatrix}
\sim
\mathcal N
\left(
\begin{pmatrix}
M_o\\
R_o (M_o)
\end{pmatrix},
\Sigma_o
\right),
\end{equation}
where $(M_o,R_o(M_o))$ denotes the noiseless $M(R)$ pair obtained from the TOV solution for the selected EoS and 
\begin{equation}
\Sigma_o=
\begin{pmatrix}
\sigma_{M,o}^2 &
\rho_o\sigma_{M,o}\sigma_{R,o}\\
\rho_o\sigma_{M,o}\sigma_{R,o} &
\sigma_{R,o}^2
\end{pmatrix}.
\end{equation}
The uncertainties are independently sampled for each observation according to
\begin{eqnarray}
\sigma_{M,o}&\sim&\mathcal U(0.05,0.10)\,M_\odot,\nonumber\\ 
\sigma_{R,o}&\sim&\mathcal U(0.10,0.30)\,\mathrm{km},\\ 
\rho_o&\sim&\mathcal U(-0.5,0.5).\nonumber
\end{eqnarray} 
Each observation is represented by \(N_S\) samples drawn independently from the multivariate Gaussian distribution defined above. These samples provide a simplified representation of the observational uncertainty associated with each measurement, rather than a single point estimate. In real astrophysical analyses, observations are commonly represented by posterior distributions over the measured quantities, which are often provided as publicly available sets of posterior samples in the \((M,R)\) plane. Consequently, the inference model is designed to operate directly on sets of samples while accommodating an arbitrary number of observations.

Unless otherwise stated, we adopt $N_S=300$ synthetic samples per observation, while the total number of observations, $N_{obs}$, is randomly selected between $5$ and $40$ during training.
Although the architecture can accommodate a variable number of samples per observation, we fix \(N_S\) during training so that all synthetic observations are represented by the same number of samples and therefore receive equal treatment by the network.

We acknowledge that this Gaussian construction provides a deliberately simplified representation of observational uncertainties. It is intended to generate controlled synthetic samples rather than to reproduce the full complexity of current observational posteriors. 

\subsection{Variable-size observational sets}
The resulting dataset has a hierarchical structure,
\[
\text{EoS}
\longrightarrow
\text{observations}
\longrightarrow
\text{samples},
\]
where each EoS is associated with a variable number of observations and each observation contains a fixed number of synthetic samples. This representation closely resembles the output of astrophysical analyses
and naturally motivates the hierarchical structure of the proposed
inference framework.

Each training example corresponds to a single EoS and is associated with a
variable number of mock observations. We denote the collection of observations
for the $i$-th EoS by $\mathcal{O}_i
=
\left\{
O_{i,j}
\right\}_{j=1}^{N_{obs,i}},$ where $i$ labels the EoS, $N_{obs,i}$ is the number of observations generated for that EoS, and $O_{i,j}$ denotes the $j$-th observation. Each observation is represented by a set of $N_S$ measurement samples,
\begin{equation} 
O_{i,j}
=
\left\{
x_{i,j,s}
\right\}_{s=1}^{N_S},
\qquad
x_{i,j,s}\in\mathbb{R}^{d_{\rm in}}\,,
\end{equation}
where $s$ labels synthetic measurement samples associated with the $j$-th observation
and $d_{\rm in}=2$ corresponds to the mass and radius of an individual sample.

For efficient mini-batch training, all EoSs within a batch are padded to the largest number of observations, $N_{\max}=\max_i N_{obs,i}$, yielding an input tensor of shape $(B,N_{\max},N_S,d_{\rm in}),$ where $B$ denotes the batch size. A binary mask is stored alongside the tensor
to distinguish valid observations from padded elements during the pooling
operations.
This representation naturally accommodates a variable number of observations
while preserving the posterior information associated with each measurement.

\section{The model}
\label{model}
NS-UNO follows the SBI NPE approach, in which a neural density estimator is trained to approximate the posterior distribution of the EoS parameters conditioned on observational data. The framework consists of two main components: a hierarchical DeepSets architecture \cite{zaheer2018deepsets} that maps a variable number of observations, each represented by a set of measurement samples, into a fixed-dimensional context vector, and a CNF that learns the posterior distribution of the EoS conditioned on this representation.

Conventional neural networks require fixed size inputs and are therefore not directly applicable to our problem, since the number of observations varies from one EoS to another. Furthermore, the ordering of both the observations and the posterior samples within each observation is arbitrary and should not affect the inferred posterior. Consequently, the
architecture must satisfy two key requirements: it must naturally accommodate variable sized inputs and remain permutation-invariant for both observation and sample sets.

\subsection{Hierarchical DeepSets}

DeepSets~\cite{zaheer2018deepsets} provide a natural solution to these
requirements. They map an unordered set of arbitrary size into a
fixed dimensional representation while guaranteeing permutation invariance.
According to the DeepSets representation theorem, any permutation-invariant
function can be written as
\begin{equation} 
f(X)=
\rho
\left(
\operatorname{pool}_{x\in X}
\phi(x)
\right)\,,
\end{equation}
where $\phi$ is an element-wise embedding network, $\operatorname{pool}$ is a
permutation-invariant aggregation operator, and $\rho$ maps the aggregated
representation into the final set representation. The pooling operator summarizes all elements of the set into a single fixed-size
vector. Typical choices include the mean, sum, second moment or maximum over the embedded elements. Our input contains a two-level hierarchy: each EoS is represented by a set of
observations, while each observation is itself represented by a set of
posterior samples. We therefore apply the DeepSets construction
hierarchically~\cite{lee2019set}. 

\medskip
\noindent\textbf{Level 1: within-observation pooling.} Each observation $O_{i,j}$ is represented by a set of $N_S$ measurement samples, $O_{i,j}=\{x_{i,j,s}\}_{s=1}^{N_S},$
where $x_{i,j,s}\in\mathbb{R}^{d_{\rm in}}$ denotes the $s$-th measurement sample
of the $j$-th observation associated with the $i$-th EoS, and
$m_{i,j,s}\in\{0,1\}$ is the corresponding validity mask.

Each sample is first embedded through a shared multilayer perceptron (MLP)
$\phi_1:\mathbb{R}^{d_{\rm in}}\rightarrow\mathbb{R}^{d_h},$
which is applied independently to every measurement sample, yielding
$z_{i,j,s}=\phi_1(x_{i,j,s})$.
To obtain a permutation-invariant representation of each observation, the
embedded samples are aggregated using masked mean pooling together with the
masked second moment,
\begin{equation}
\mu_{i,j}
=
\frac{\sum_s m_{i,j,s}\,z_{i,j,s}}
{\sum_s m_{i,j,s}}
\ \ \ \text{and}\ \ \   
\nu_{i,j}
=
\frac{\sum_s m_{i,j,s}\,z_{i,j,s}^{\odot2}}
{\sum_s m_{i,j,s}}\,,
\end{equation}
where $\odot2$ denotes element-wise squaring. The concatenation
$[\mu_{i,j};\nu_{i,j}]$ captures both the first- and second-order statistics of the posterior samples, producing a richer representation than mean pooling alone while remaining permutation-invariant.
A second network, $\rho_1:\mathbb{R}^{2d_h}\rightarrow\mathbb{R}^{d_{\rm obs}},$ maps the pooled representation to an observation embedding,
$h_{i,j}=\rho_1([\mu_{i,j};\nu_{i,j}]).$ 

\medskip
\noindent\textbf{Level 2: observations aggregation.} The observation embeddings are first projected into a higher-dimensional
space through a second shared network, $\phi_2:\mathbb{R}^{d_{\rm obs}}
\rightarrow
\mathbb{R}^{d_{h_2}},$
yielding $u_{i,j}=\phi_2(h_{i,j}).$

Rather than assigning the same importance to every observation, we aggregate
the observation embeddings using an attention pooling mechanism
\cite{ilse2018attention,lee2019set}. An attention network $a:\mathbb{R}^{d_{h_2}} \rightarrow \mathbb{R}$, 
assigns a scalar score to each observation, which is converted into normalized attention weights through a \texttt{softmax} function,
\begin{equation}
\alpha_{i,j}
=
\frac{\exp(a(u_{i,j}))}
{\sum_{k=1}^{N_{obs,i}}\exp(a(u_{i,k}))}\,.
\end{equation}
The resulting context representation for the $i$-th EoS is obtained as the
weighted average
\begin{equation}
s_i
=
\sum_{j=1}^{N_{obs,i}}
\alpha_{i,j}\,
u_{i,j}\,,
\end{equation}
which remains permutation-invariant because the attention weights depend only on the observation embeddings. Finally, a network 
$\rho_2: \mathbb{R}^{d_{h_2}} \rightarrow \mathbb{R}^{d_{\rm ctx}},$ maps the pooled representation into the context vector, 
$c_i = \rho_2(s_i),$ which conditions the CNF.

\medskip
\noindent\textbf{Tensor shapes.} Denoting the batch size by $B$, the maximum number of observations in the
batch by $N_{\max}$, and the number of measurement samples per observation by
$N_S$, the architecture processes tensors according to the following scheme: 
\begin{align*}
&(B,N_{\max},N_S,d_{\rm in})
\xrightarrow{\phi_1}
(B,N_{\max},N_S,d_h) \xrightarrow{}\\
&\xrightarrow{\text{moment pooling}}
(B,N_{\max},2d_h)
\xrightarrow{\rho_1}
(B,N_{\max},d_{\rm obs}) \xrightarrow{}\\
&\xrightarrow{\phi_2}
(B,N_{\max},d_{h_2})
\xrightarrow{\text{attention}}
(B,d_{h_2})
\xrightarrow{\rho_2}
(B,d_{\rm ctx}).
\end{align*}

The hierarchical DeepSets architecture is particularly well suited to this problem,
as it naturally accommodates a variable number of observations while remaining
permutation-invariant. Since neither the ordering of the observations nor that of the samples within each observation carries physical meaning, the inferred posterior should be independent of their ordering.
The hierarchical pooling operations ensure this property while mapping an arbitrary sized observational set into a fixed-dimensional context vector. This context vector therefore provides a learned summary of all available observations and is subsequently used to condition the CNF. 
\begin{figure*}[!hbt]
    \centering
    \includegraphics[width=1\linewidth]{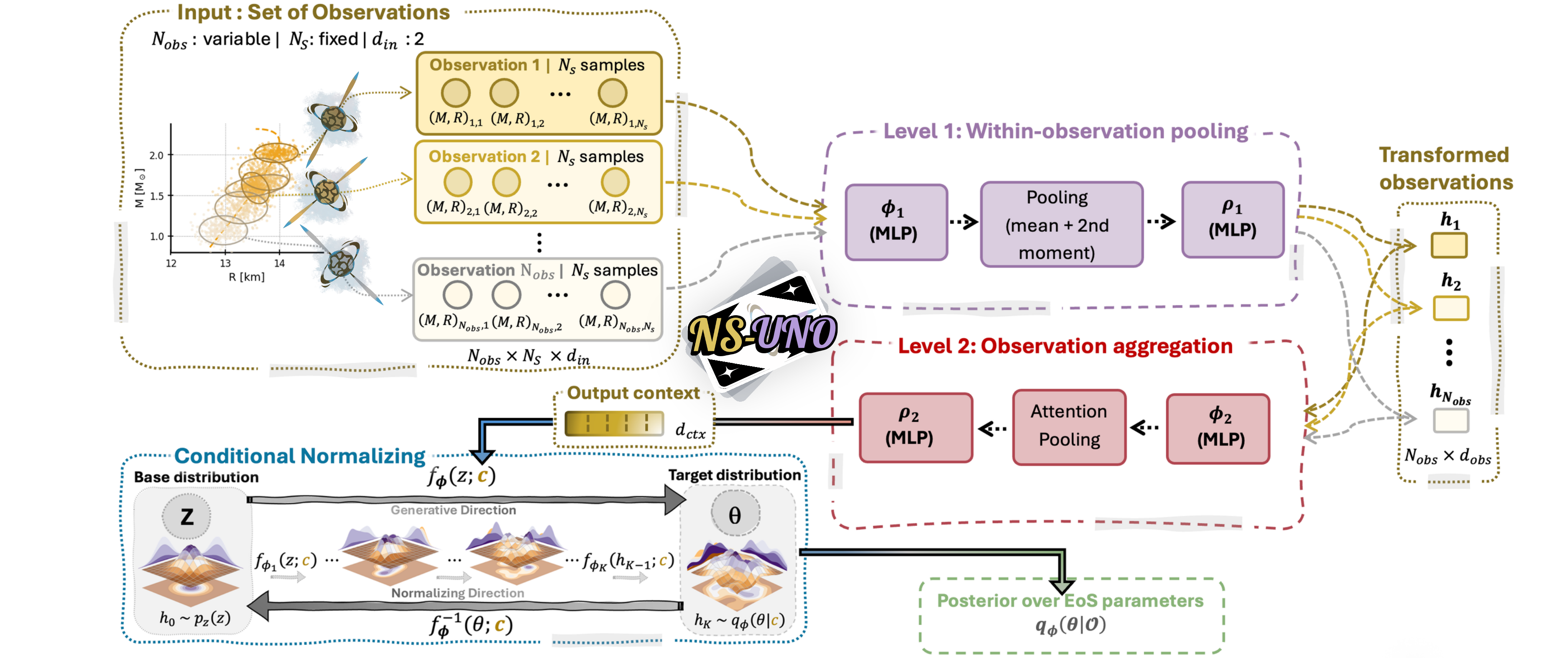}
\caption{
Schematic representation of the NS-UNO architecture for one EoS. Each of the \(N_{\rm obs}\) observations is represented by \(N_S\) synthetic measurement samples, which are first embedded and aggregated using mean and second moment pooling (Level~1). The resulting observation representations are then combined through attention pooling (Level~2) to produce a fixed-dimensional context vector that conditions the CNF, and then allows obtaining the posterior over the EoS parameters. 
}
    \label{fig:scheme}
\end{figure*}

\medskip

\noindent\textbf{Hyperparameters.} All MLP use two hidden layers with ReLU activations. The
architecture dimensions are
$d_{\rm in}=2$,
$d_h=128$,
$d_{\rm obs}=128$,
$d_{h_2}=256$, and
$d_{\rm ctx}=128$.

\subsection{Conditional Normalizing Flows}

The context vector $c_i$ produced by the hierarchical DeepSets model
provides a fixed-dimensional summary of the complete observational set
$\mathcal{O}_i$ associated with the $i$-th EoS. This representation is then
used to condition a CNF, which approximates the
posterior distribution of the EoS parameters $\theta$. In our implementation,
\begin{equation}
\theta=
\left[
\log p(n_1),\,
\log p(n_2),\,
\ldots,\,
\log p(n_{20})
\right]\,,
\end{equation} 
where the pressure $p$ is evaluated at the 20 fixed baryon-density points $n_i$, $i\in (1,\dots,20)$ introduced in Sec.~\ref{dataset}. Following our previous work \cite{carvalho2025neural}, the CNF transforms a simple base distribution $p_z(z)$ into the target posterior through a sequence of invertible transformations $f_\phi$,
\begin{equation}
q_\phi(\theta \mid c_i)
=
p_z
\!\left(
f_\phi^{-1}(\theta;c_i)
\right)
\left|
\det
J^{-1}_{f_\phi}
\right|.
\end{equation}
The flow is implemented using rational quadratic spline coupling transforms,
in which each coupling layer updates a subset of the variables conditioned on
the remaining ones. This construction guarantees invertibility while allowing
the Jacobian determinant to be evaluated efficiently.

Training minimizes the expected negative log-posterior density of the reference EoS parameters under the learned posterior approximation,
\begin{equation}
\mathcal{L}_{\rm NLL}
=
-
\mathbb{E}_{(\theta,c)}
\left[
\log q_\phi(\theta\mid c)
\right].
\end{equation}

To enforce physically consistent EoS, we include a
monotonicity regularization term that penalizes decreasing pressure with
increasing baryon density,

\begin{equation}
\mathcal{L}_{\rm mono}
=
\sum_{k=1}^{19}
\max
\left(
0,\,
p(n_k)-p(n_{k+1})
\right).
\end{equation}

The total loss is therefore
\begin{equation}
\mathcal{L}
=
\mathcal{L}_{\rm NLL}
+
\lambda\,
\mathcal{L}_{\rm mono}\,,
\label{eq:total_loss} 
\end{equation}
where $\lambda$ controls the strength of the physical regularization. A schematic representation of the model is shown in Fig.~\ref{fig:scheme}. 
Implementation details, including the network architecture and optimization
strategy, are provided in Appendix~\ref{ssec:implementation}.

\section{Results}
\label{results}

We evaluate the proposed NS-UNO framework through a series of complementary experiments. We first assess the reconstruction accuracy and calibration of the inferred posterior over the complete test set. We then investigate the impact of the (unknown) NS mass distribution and the  number of available observations, the robustness to observational uncertainties, and the ability of the model to generalize across different EoS families. Finally, we demonstrate the framework in a qualitative application to current NICER and the LVK GW170817 observations.

\subsection{Reconstruction accuracy}

\begin{figure}[!hbt]
    \centering
    \includegraphics[width=1\linewidth]{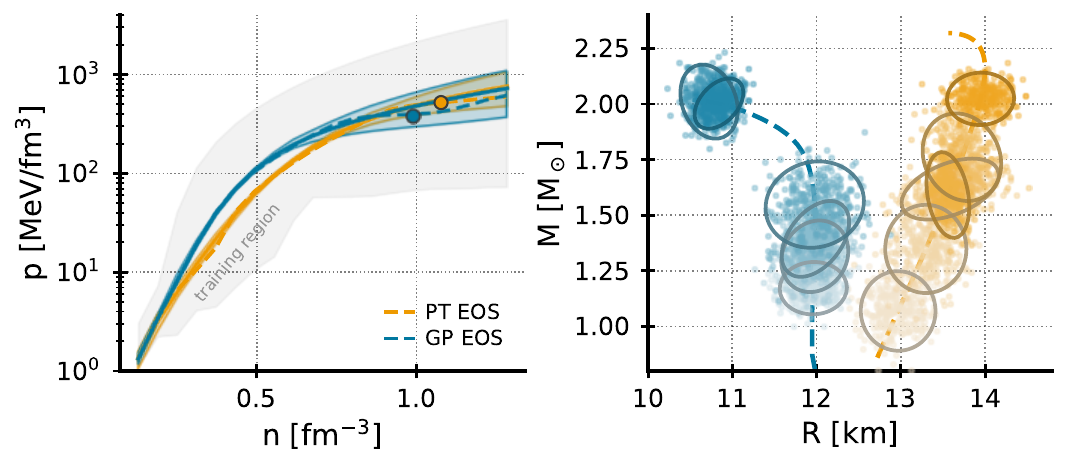}
    \caption{
    Left panel: representative posterior reconstructions for one randomly selected GP EoS
    (blue color) and one PT EoS (yellow color). The dashed curves denote the true
    EoS, while the solid line curves and shaded regions show the posterior
    median and the corresponding 90\% credible interval, and the dots $n_{c,max}$. The grey region denotes the
    pressure range covered during training. Right panel:  corresponding simulated $M(R)$ observations (dots) used to condition the reconstruction, and the ``ground truth'' TOV solutions (dashed curves).}
    \label{fig:1}
\end{figure}

We first study the model by reconstructing one representative EoS from both the PT and GP in the test set. Figure~\ref{fig:1}
shows one randomly selected EoS from each dataset together with the corresponding posterior reconstruction. For consistency, we use the same number of observations, $N_{obs}=6$, for both EoSs.

Through all obtained samples we impose monotonicity, thermodynamic consistency and causality as hard physical constraints. For the pQCD matching criterion we adopt the least restrictive prescription where we use the QCD likelihood function implementation \cite{komoltsev_2023_7781233}.

For both examples, the posterior median accurately reproduces the true EoS over the density range directly constrained by the observations. As expected, the posterior uncertainty gradually increases at higher baryon densities, reflecting the reduced amount of observational information available beyond the maximum central density $n_{c,max}$, represented by a dot in each EoS, reached by the observed NS. The right plot shows the corresponding mock $M(R)$ observations used to condition the inference. This illustrates how the observational mass distribution determines the region of the EoS that is directly constrained.

The examples above provide an illustrative view of the reconstruction behaviour for individual EoSs. To assess the reconstruction accuracy systematically across the two EoS families, we next evaluate the model on a larger and balanced sample of test EoSs. Next we  estimate  the reconstruction accuracy for both datasets. Although the GP and PT datasets are not equally represented during training, all results presented in this section are computed using 4\,000 randomly selected EoSs from each test set, providing a balanced comparison between the two EoS families. For each EoS, we draw $N_p=1\,000$ samples from the inferred posterior distribution, yielding posterior realizations $\theta_P^{(i,l)}(n)$, where $i=1,\ldots,4\,000$ labels the EoS and $l=1,\ldots,N_p$ labels the posterior samples predicted by the model. 
Each EoS is associated with a set of observations \(O_i\) containing a variable number of observations, \(N_{\rm obs,i}\), within the range considered during training. Section~\ref{sec:Mock_observations} contains details on the construction of the mock observations. All reconstruction metrics are computed from these posterior samples.

To quantify the reconstruction accuracy, we compute the relative residual for each EoS,
\begin{equation}
\label{eq:RREs}
\mathrm{RelRes}^{(i)}(n)=
\mathrm{Med}_l
\left[
\frac{\theta_P^{(i,l)}(n)-\theta_T^{(i)}(n)}
{\theta_T^{(i)}(n)}
\right]\times100,
\end{equation}
where the median ($\mathrm{Med}$) is taken over the $N_p= 1\,000$ posterior samples and $\theta_T^{(i)}(n)$ denotes the corresponding ground truth pressure profile. The reconstruction error is then summarized over the $4\,000$ test EoS through the median absolute relative residual, 
\begin{equation}
\label{reselid}
|\overline{\mathrm{RelRes}}(n)|=
\mathrm{Med}_i
\left|
\mathrm{RelRes}^{(i)}(n)
\right|,
\end{equation}
which provides a robust estimate of the typical reconstruction error as a function of baryon density.

\begin{figure}[!hbt]
    \centering
    \includegraphics[width=0.49\linewidth]{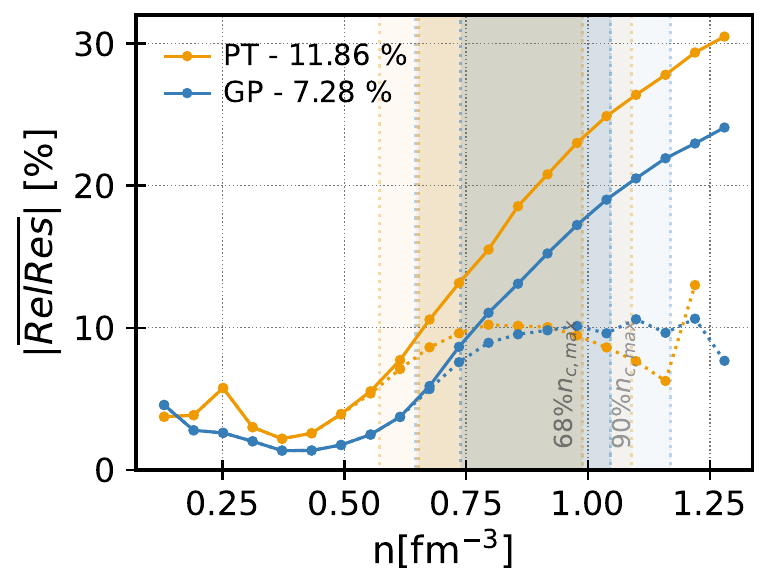} 
    \includegraphics[width=0.49\linewidth]{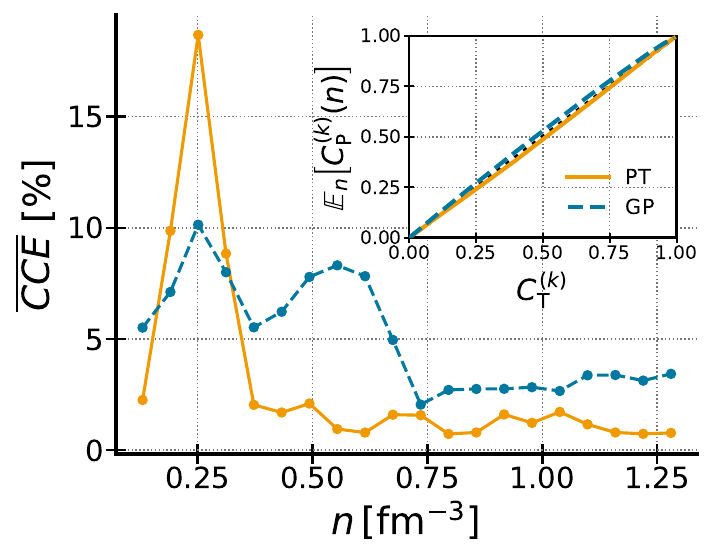}
\caption{
Left panel: Median absolute relative residual as a function of baryon density for the PT (yellow) and GP (blue) test sets. Solid curves show the reconstruction error over the full density range, while dotted curves are computed only up to the maximum central density, $n_{c,\rm max}$, reached by the observed NSs. The vertical shaded bands indicate the 68\% (dark) and 90\% (light) intervals of the $n_{c,\rm max}$ distribution across the dataset.
Right panel: Coverage Calibration Error (CCE) as a function of baryon density for the PT and GP test sets, Eq. \ref{Eq:CCE}. The inset shows the calibration curve obtained by averaging the empirical coverage over all density points and plotting it against the corresponding credible levels. A perfectly calibrated posterior lies on the diagonal.
}
    \label{fig:median_cce}
\end{figure}

Left plot of Fig.~\ref{fig:median_cce} presents the reconstruction error for the PT (yellow curves) and GP (blue curves) test sets. Over the full density range, the model achieves a median absolute relative residual of $7.28\%$ for the GP dataset and $11.86\%$ for the PT dataset. Compared to our previous model in \cite{carvalho2025neural}, we obtain overall better accuracy. At low densities, both datasets exhibit comparable reconstruction accuracy. However, the residual gradually increases above $n\,{\simeq}\,0.6~\mathrm{fm}^{-3}$, particularly for the PT models, reflecting the larger diversity of high density EoSs contained in the PT dataset. 

The dotted curves show the reconstruction error computed only up to the maximum central density reached by the available observations. In both datasets, the filtered residual remains significantly lower than the full residual, demonstrating that the model reconstructs the EoS most accurately within the density regime directly constrained by the observed NSs. Beyond this region, the posterior naturally broadens as the model infers the
EoS in density regimes that are not directly constrained by the available observations.

As a SBI method, NPE requires not only accurate posterior predictions but also well-calibrated uncertainty estimates. To assess the calibration of the inferred posterior, we compute the relative Coverage Calibration Error (CCE) at each baryon density,
\begin{equation}
\label{Eq:CCE}
\overline{\mathrm{CCE}} (n)=
\frac{1}{K}
\sum_{k=1}^{K}
\left|
\frac{
C_{\mathrm{P}}^{(k)}(n)-C_{\mathrm{T}}^{(k)}
}{
C_{\mathrm{T}}^{(k)}
}
\right|
\times100,
\end{equation}
 the average is computed over the $K$ credible levels,
$C_{\mathrm{T}}^{(k)}\in[0,1]$.
Here,
$C_{\mathrm{P}}^{(k)}(n)\in[0,1]$
denotes the empirical coverage at baryon density $n$, i.e. the fraction of test EoSs whose true pressure at density $n$ lies within the nominal credible interval corresponding to the $k$-th credible level.

Right panel of Fig.~\ref{fig:median_cce} shows
$\overline{\mathrm{CCE}}(n)$
as a function of baryon density for both datasets, while the inset displays the calibration curve obtained by averaging the empirical coverage over all density points,
$\mathbb{E}_n[C_{\mathrm{P}}^{(k)}(n)]$,
and plotting it against the corresponding nominal credible level. 

The calibration curves closely follow the diagonal, indicating excellent agreement between the empirical and nominal coverages. Consistently, $\overline{\mathrm{CCE}}$ remains below approximately $20\%$ over the entire density range and is typically below $10\%$. The largest deviations occur at the lowest densities, where the posterior is most constrained and therefore more sensitive to small discrepancies between the true and predicted confidence intervals. Overall, these results demonstrate that the proposed framework provides well-calibrated posterior uncertainty estimates in addition to accurate EoS reconstructions.

\subsection{Effects of observational coverage} \label{Mass_coverage}

\begin{figure}[!hbt]
    \centering
    \includegraphics[width=1\linewidth]{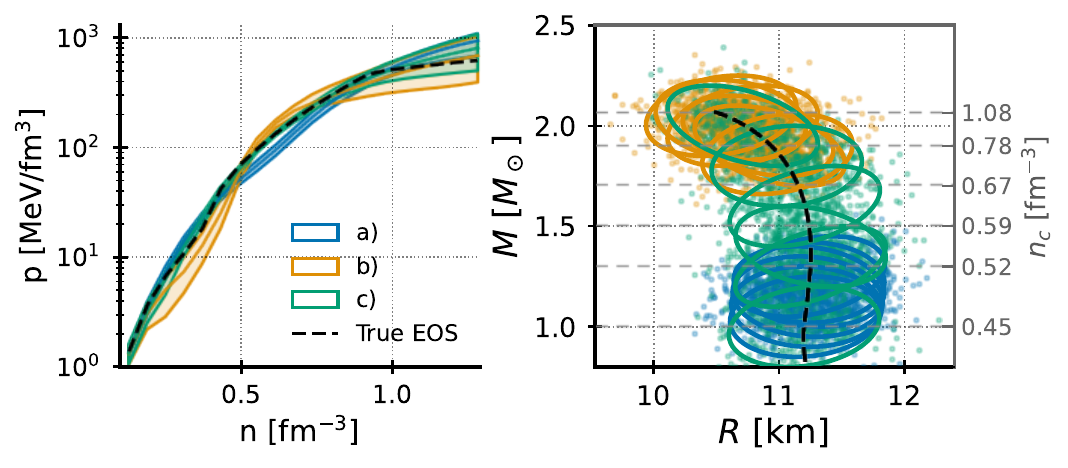}
\caption{
Reconstruction of a representative EoS from the test set using three different observational mass distributions:
a) six low-mass observations (${\sim}1.1\,M_\odot$),
b) six high-mass observations (${\sim}2.0\,M_\odot$) and
c) six observations approximately uniformly distributed over the full TOV sequence.
Left: reconstructed EoSs with their 90\% credible intervals, with the true curve shown in dashed black.
Right: corresponding $M(R)$ observations. The secondary vertical axis indicates the central densities $n_c$ of the ``ground truth'' TOV curves. 
}
    \label{fig:mass_interval}
\end{figure}

To investigate how the observational NS mass range affects the reconstruction, Fig.~\ref{fig:mass_interval} shows the prediction for a representative EoS from the test set using three different observational configurations:
a) six observations clustered around low-mass NSs (${\sim}1.1\,M_\odot$),
b) six observations clustered around high-mass NSs (${\sim}2.0\,M_\odot$),
and c) six observations approximately uniformly distributed along the entire TOV curve.

The left plot compares the reconstructed EoSs, while the right plot shows the corresponding observational samples in the $M(R)$ plane together with the central densities they probe. Since each NS mass is associated with a different central density, the observations directly constrain only a limited region of the EoS. Consequently, when only low-mass observations are available, case a), the reconstruction is tightly constrained only up to the central densities reached by those NSs, and the posterior uncertainty increases rapidly at higher densities. The opposite behaviour is observed for case b), where only the high-density region is strongly constrained. The most accurate and stable reconstruction is obtained when observations span the entire TOV sequence, case c). In this case, the inferred posterior remains consistent with the true EoS over the full density range considered.

\begin{figure}[!hbt]
    \centering
    \includegraphics[width=1\linewidth]{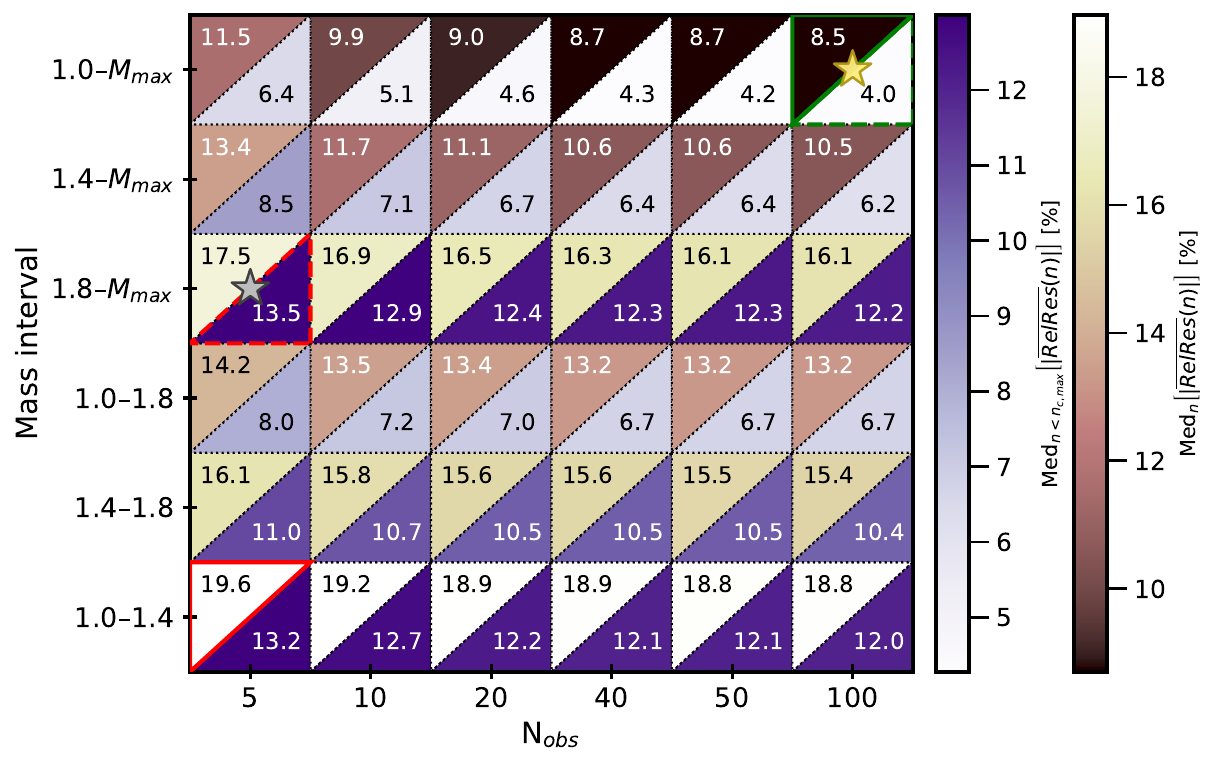}
\caption{
Median absolute relative residual (\%) as a function of the number of observations, $N_{\rm obs}$, and the observational mass interval, evaluated using 1000 randomly selected test EoSs. Six mass intervals are considered:
$[1.0,1.4]$,
$[1.4,1.8]$,
$[1.0,1.8]$,
$[1.8,M_{\rm max}]$,
$[1.4,M_{\rm max}]$, and
$[1.0,M_{\rm max}]\,M_\odot$.
Each cell is divided diagonally: the lower (purple) triangle shows
$\mathrm{Med}_{n<n_{c,\rm max}}\!\left[|\overline{\mathrm{RelRes}}(n)|\right]$,
computed only up to the maximum central density reached by the observations, while the upper(brown) triangle shows
$\mathrm{Med}_{n}\!\left[|\overline{\mathrm{RelRes}}(n)|\right]$,
computed over the full density range. The green and red triangles identify the best and worst reconstruction errors, respectively, while the gold and grey stars mark the largest and smallest differences between the filtered and full-density errors.
}
\label{fig:MAPE_diff_mass}
\end{figure}

Figure~\ref{fig:MAPE_diff_mass} investigates how the reconstruction accuracy
depends on both the number of available observations and the mass range they
cover. We randomly selected 1000 EoSs from the test set and considered six
observational mass intervals:
$[1.0,1.4]$,
$[1.4,1.8]$,
$[1.0,1.8]$,
$[1.8,M_{\rm max}]$,
$[1.4,M_{\rm max}]$, and
$[1.0,M_{\rm max}]$, in units of $M_\odot$.
For each, the model was evaluated using $N_{\rm obs}=5,10,20,40,50$, and $100$ observations.

Each heatmap cell is divided diagonally to compare the reconstruction error computed over two density ranges. The lower (purple) triangle reports the median absolute relative residual evaluated only up to the maximum central density reached by the observations (``filtered'') $\mathrm{Med}_{n<n_{c,\rm max}}\!\left[|\overline{\mathrm{RelRes}}(n)|\right]$, while the upper (brown) triangle corresponds to the error computed over the full density range $\mathrm{Med}_{n}\!\left[|\overline{\mathrm{RelRes}}(n)|\right]$.

Several trends emerge. Firstly, increasing the number of observations consistently improves the reconstruction accuracy, although the improvement becomes progressively smaller beyond approximately
$N_{\rm obs}\simeq40$. Secondly, for every observational configuration, the filtered reconstruction error remains substantially lower than the error computed over the full density range, confirming that the model performs best within the density ranges constrained by the available observations.

This distinction is particularly important from a theoretical perspective. Since NS observations can only probe densities up to the maximum central density reached by the observed stars, no inference framework can be expected to accurately reconstruct the EoS beyond this regime. The filtered metric therefore quantifies the predictive performance within the physically accessible density range, whereas the full density metric additionally reflects the model extrapolation to unconstrained densities.

More importantly, the mass coverage of the observations has a considerably
larger impact on the reconstruction accuracy than the number of observations
alone. The smallest reconstruction errors are obtained when the observations
span the full accessible mass range,
$[1.0,M_{\rm max}]\,M_\odot$, whereas restricting the observations to either
only low-mass NSs, $[1.0,1.4]\,M_\odot$, or only high-mass NSs, $[1.8,M_{\rm max}]\,M_\odot$, leads to significantly poorer performance.
This behaviour indicates that observations sampling a broad range of NS masses provide complementary constraints on different density regions of the EoS.

The contrast between the filtered and full-density reconstruction is largest
for the $[1.0,M_{\rm max}]\,M_\odot$ interval with
$N_{\rm obs}=100$, highlighted by the gold star in
Fig.~\ref{fig:MAPE_diff_mass}. In this case, the observations constrain only
part of the density range, allowing the model to reconstruct the directly
probed region with high accuracy while naturally increasing its uncertainty at
higher densities. Conversely, the smallest difference is found for the
$[1.8,M_{\rm max}]\,M_\odot$ interval with $N_{\rm obs}=5$ (grey star),
where the observations already probe relatively high densities, reducing the
gap between the filtered and full-density evaluations.

Overall, these results demonstrate that extending the mass range covered by
the observations is more beneficial than simply increasing their number.
Including high mass NS substantially improves the reconstruction of
the high density EoS, whereas adding many observations confined to a narrow
mass interval provides comparatively limited additional information.

\subsection{Generalization to different EoS families}\label{EoS_families}

To evaluate the NS-UNO ability to generalize beyond the EoS ensembles used during training, we consider twelve representative EoSs drawn from different theoretical models.

\begin{figure}[!hbt]
    \centering
    \includegraphics[width=0.9\linewidth]{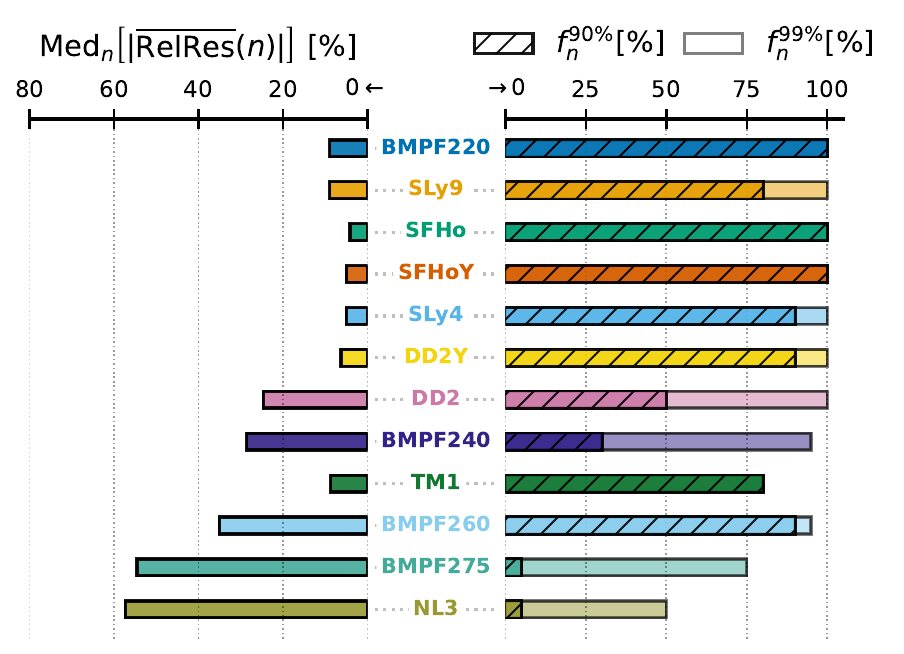}
\caption{
Generalization performance of the proposed framework on representative EoSs belonging to theoretical families not included in the training ensembles. The left panel shows the median reconstruction error,
$\mathrm{Med}_n\!\left[|\overline{\mathrm{RelRes}}(n)|\right]$.
The right panel shows $f_n^{90\%}$ (hatched bars) and $f_n^{99\%}$ (lighter color bars), defined as the percentage of density points contained within the predicted 90\% and 99\% credible intervals, respectively. Smaller values of the reconstruction error and larger empirical coverages correspond to better reconstruction quality.}
    \label{fig:relres_other}
\end{figure}

Figure~\ref{fig:relres_other} summarizes the reconstruction quality for each EoS through two complementary quantities. The left panel shows the median absolute relative residual, $\mathrm{Med}_n\!\left[|\overline{\mathrm{RelRes}}(n)|\right]$, defined in Eq.~\ref{reselid}, which quantifies the typical relative residual of the inferred EoS, with smaller values corresponding to more accurate predictions. The right panel shows $f_n^{90\%}$ and $f_n^{99\%}$, defined as the fraction of density points whose true pressure lies within the predicted 90\% and 99\% credible intervals, respectively. Larger values of $f_n^{x}$ indicate that the inferred posterior captures the true EoS over a larger fraction of the density range.

The present test considers the following EoSs: SLy4 and SLy9, which were developed to reproduce the properties of nuclear matter in neutron-rich environments \cite{Chabanat:1995tea,Chabanat:1997un}; the relativistic mean field (RMF) models BMPF220, BMPF240, BMPF260, and BMPF275, which differ in the incompressibility at nuclear saturation density \cite{malik2023spanning}; DD2, a relatively stiff density-dependent RMF EoS \cite{Typel:2009sy}; SFHo, a softer RMF EoS \cite{Steiner:2012rk}; and the RMF models NL3 \cite{Lalazissis:1996rd} and TM1 \cite{Sugahara:1993wz}, both fitted to finite nuclei, with TM1 exhibiting a softer behaviour at high densities. We additionally consider the hyperonic extensions DD2Y and SFHoY \cite{fortin2018hyperons}, which include hyperons as additional degrees of freedom and therefore provide a more stringent test of the model's ability to generalize to EoSs with different microscopic compositions.

Overall, most of the considered EoSs exhibit median absolute relative residual below approximately $40\%$, while simultaneously capturing a large fraction of the density points within both the predicted 90\% and $99\%$ credible intervals. In particular, SFHo achieves the highest value of $f_n^{90\%}$, with all density points lying inside the predicted $90\%$ credible interval while also exhibiting one of the smallest reconstruction errors. Similarly, BMPF220, SLy4, and SLy9 EoS models combine low reconstruction errors with high values of both $f_n^{90\%}$ and $f_n^{99\%}$, indicating that the inferred posterior accurately recovers the EoS over most of the considered density range.

The reconstruction quality nevertheless varies across EoS families. TM1 and BMPF240 models maintain $f_n^{90\%}$ values above $80\%$, despite exhibiting moderately larger reconstruction errors than the best performing models. DD2 displays one of the lowest reconstruction errors but only approximately half of the density points are contained within the predicted $90\%$ credible interval, whereas all density points lie inside the predicted $99\%$ credible interval. This indicates that the inferred posterior remains informative but that the $90\%$ credible interval is comparatively narrow for this EoS. Conversely, BMPF260 exhibits a larger reconstruction error while still retaining high values of both $f_n^{90\%}$ and $f_n^{99\%}$, indicating that the inferred uncertainties remain well calibrated despite the reduced reconstruction accuracy.

The largest reconstruction errors are obtained for NL3 and BMPF275, which also exhibit the lowest values of both $f_n^{90\%}$ and $f_n^{99\%}$. Both models display relatively extreme $M(R)$ relations compared with the remaining EoSs considered here, suggesting that they occupy regions of the EoS space that are more challenging for the proposed framework to reconstruct accurately.

The hyperonic models DD2Y and SFHoY exhibit reconstruction errors comparable to those of their nucleonic counterparts while retaining $f_n^{90\%}$ values above $80\%$ and nearly complete coverage at the 99\% credible level. More generally, all the EoSs considered in this section belong to theoretical families that are distinct from the agnostic PT and GP ensembles used for training. The ability of the proposed framework to accurately reconstruct these previously unseen models demonstrates that it captures the underlying relationship between NS observables and the EoS, enabling robust generalization across a broad range of theoretical descriptions of dense matter.

EoSs containing phase transitions are not included in the present generalization analysis, as such features are not explicitly represented in the training data and were not reconstructed reliably by the current implementation. Nevertheless, a proof of concept study investigating the model sensitivity to phase transition like signatures in the $M(R)$ relation is presented in Appendix~\ref{app:phaseTransition}.

\subsection{Generalization to noisier and real observations}

The observational uncertainties adopted during training are intentionally simplified and do not reproduce the full complexity of current NICER and the LVK posteriors. They are instead intended to approximate the precision expected from next-generation multimessenger observations. Before applying the model to real observations, we therefore investigate whether the learned posterior responds consistently to increasing measurement uncertainty.

For each test EoS we generate synthetic observations with progressively larger uncertainties and compute
\begin{eqnarray}\label{Eq_noise}
&&\Delta \left[\log q_{\phi}(\theta_{\rm T}\mid \mathcal{O}_{\Delta\sigma}) \right] = \nonumber\\ 
&& = \log q_\phi(\theta_{\rm T}\mid \mathcal{O}_{\Delta\sigma=1})-\log q_\phi(\theta_{\rm T}\mid \mathcal{O}_{\Delta\sigma})\,,
\end{eqnarray}
where $\mathcal{O}_{\Delta\sigma}$ denotes the set of mock observations generated by scaling the training uncertainties by a factor $\Delta\sigma$, and $\theta_{\rm T}$ is the corresponding ground-truth EoS.\\

\begin{figure}[!hbt]
    \centering
\includegraphics[width=0.8\linewidth]{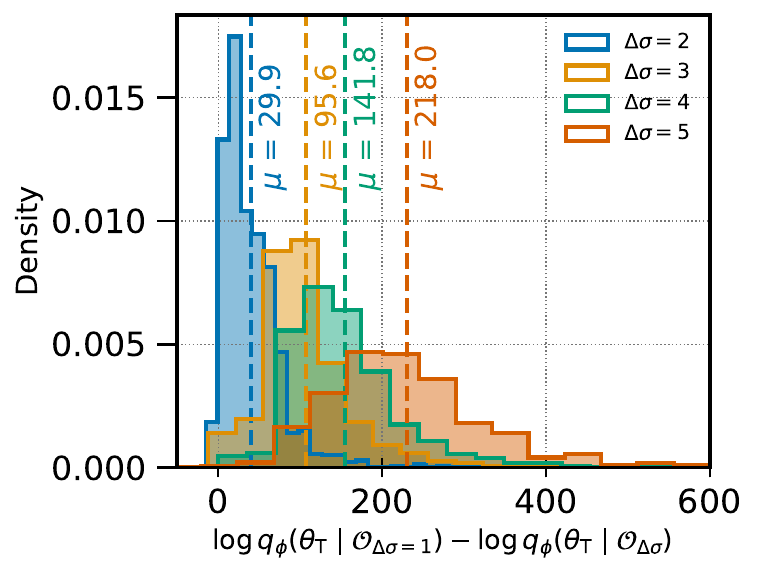}
\caption{ Distribution of the change in the log posterior density assigned by the network to the true EoS as the observational uncertainties are increased by a factor $\Delta\sigma$, relative to the trained uncertainty ($\Delta\sigma=1$). For each test EoS, we compute Eq. \ref{Eq_noise}
where $\mathcal{O}_{\Delta\sigma}$ denotes the set of mock observations generated with uncertainties scaled by $\Delta\sigma$. The dashed vertical lines indicate the mean of each distribution, with the corresponding values shown alongside the lines.
}
    \label{fig:hist_sigma}
\end{figure}

Figure~\ref{fig:hist_sigma} shows that increasing the observational uncertainty systematically reduces the log posterior density assigned to the true EoS relative to the training uncertainty.

As the observations become less informative, the posterior becomes broader and the true EoS receives a lower posterior density. This demonstrates that, despite being trained over a restricted uncertainty range, the model responds qualitatively as expected to degraded observational information.

Having established that the model behaves consistently under increasing observational uncertainty, we now perform a qualitative test using currently available NS observations. 
As the conditioning set, we use the mass and radius constraints inferred for the binary NS merger GW170817 \cite{Abbott:2018wiz,abbott2018gw170817}, together with the 
$M(R)$ measurements obtained from X-ray pulse-profile modelling by NICER. The NICER observations include PSR J0030+0451 \cite{Riley_2019}, PSR J0740+6620 \cite{salmi2024radius}, PSR J0437$-$4715 \cite{choudhury2024nicer}, PSR J1231$-$1411 \cite{salmi2024nicer}, and PSR J0614$-$3329 \cite{mauviard2025nicer}. 
For each source, we use the publicly released posterior samples. To provide a consistent input representation, each observation is represented by 2000 posterior samples. When posterior weights are available, the 2000 samples with the highest weights are retained. This choice corresponds to the maximum common sample size across the observational datasets considered.

Because the model was trained using synthetic observations with simplified Gaussian uncertainties, these data do not exactly match the uncertainty distributions of current measurements. The Gaussian uncertainties adopted during training,
$\sigma_M\in[0.05,0.10]\,M_\odot$ and
$\sigma_R\in[0.10,0.30]\,\mathrm{km}$,
are generally smaller than those of current observations, especially for the radius measurements. Moreover, the real posteriors are often non-Gaussian and exhibit irregular shapes that are not represented in the training set.

Consequently, the resulting posterior should not be interpreted as a
physical inference of the NS EoS. Instead, the purpose of this experiment is
to investigate whether the learned posterior remains sensitive to physically
meaningful differences between candidate EoSs when conditioned on real
observations.

More specifically, we ask whether the model assigns systematically different
posterior densities to representative theoretical EoSs whose $M(R)$
relations exhibit different levels of compatibility with the currently
available observations. This therefore constitutes an out of distribution
test of the framework, since both the observational uncertainties and the
posterior shapes differ from those encountered during training.

Since posterior densities are only defined up to an arbitrary normalization, their absolute values are not directly informative when comparing different candidate EoSs. We therefore report the posterior density of each candidate relative to that of the most probable EoS within the considered set,
\begin{eqnarray}
\label{Eq_obs}
&&\Delta \left[ \log q_\phi(\theta_N^{(i)}|\mathcal{O}_{real}) \right]
= \nonumber \\ 
&& = \max_j
\log q_\phi(\theta_N^{(j)}|\mathcal{O}_{real}) - \log q_\phi (\theta_N^{(i)}|\mathcal{O}_{real})\,.   
\end{eqnarray}
where $\mathcal{O}_{real}$ denotes the collection of current observations and $\theta_N^{(i)}$ denotes the i-th new candidate EoS described in Sec.~\ref{EoS_families}. 

The subtraction of the maximum posterior density, $\max_j \log q_\phi(\theta_N^{(j)}|\mathcal{O}_{real})$ simply shifts all values by a constant. Equation \ref{Eq_obs} therefore provides a relative ranking of the candidate EoSs: values closer to zero indicate posterior densities closer to that of the highest ranked candidate, while larger positive values indicate lower posterior densities relative to this reference.

\begin{figure}[!hbt]
    \centering
    \includegraphics[width=.9\linewidth]{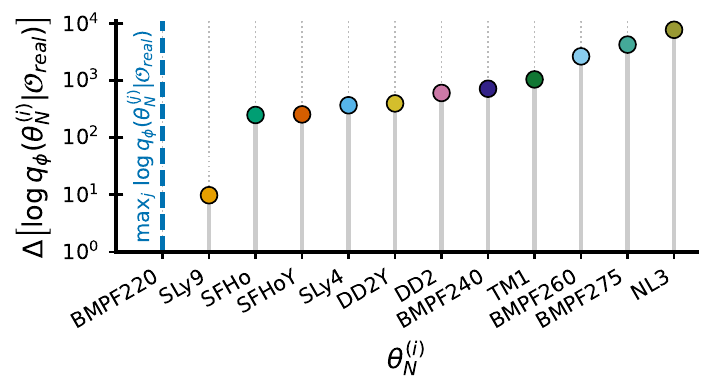}\\
    \includegraphics[width=0.8\linewidth]{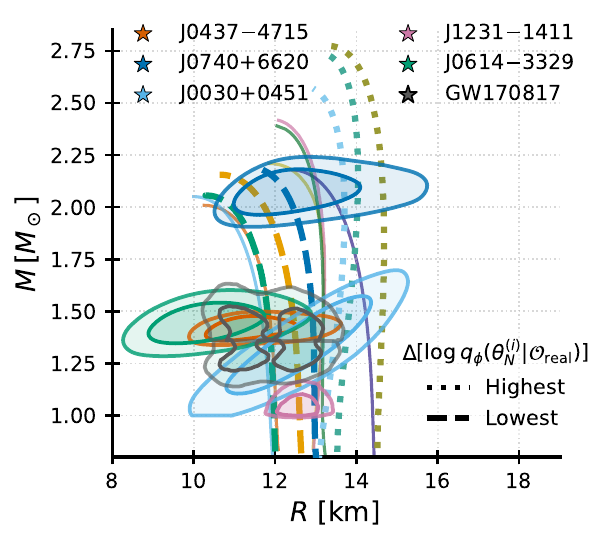}
\caption{
Upper panel: Relative posterior assigned by the trained model to a representative set of theoretical EoSs using the currently available observations described in the text. The quantity shown is $\Delta \left[ \log q_\phi(\theta_N^{(i)}|\mathcal{O}_{real}) \right]$, see Eq.~\ref{Eq_obs}, where $\mathcal{O}_{real}$ denotes the collection of current observations and $\theta_N^{(i)}$ the different EoS. Values closer to zero indicate greater compatibility with the observations according to the learned posterior, whereas increasingly positive values correspond to less favoured EoSs. The results should be interpreted only in a relative sense and do not constitute a quantitative model selection analysis. Lower panel: Corresponding $M(R)$ relations for the same EoSs, using the same colour coding as the upper plot. The dashed curves indicate the three highest ranked EoSs, while the dotted curves correspond to the three lowest ranked EoSs.}
    \label{fig:Real_obser}
\end{figure}

Figure~\ref{fig:Real_obser} summarises the results of this qualitative test, where the EoS used to test the model were the same as the ones used in Fig. \ref{fig:relres_other}. The upper plot shows the relative posterior density assigned to each candidate EoS, ordered from the highest to the lowest posterior density. The lower plot shows the corresponding $M(R)$ relations together with the currently available NICER and LVK observations, using the same colour coding as the upper plot. This allows the relative posterior ranking to be interpreted in the context of the corresponding $M(R)$ relations and the current observational constraints.

We emphasize that the quantity shown in Fig.~\ref{fig:Real_obser} is not the likelihood of the observations given a particular EoS, but rather the posterior density assigned by the learned NPE. The purpose of this
experiment is therefore not model selection, but to assess whether the
learned posterior distinguishes between candidate EoSs when presented with
real observational data.

The preferred EoSs generally pass through the regions favoured by the current observations, whereas the least preferred models exhibit larger deviations. Nevertheless, many of the $M(R)$ relations overlap within the observational uncertainties; among the EoSs considered, BMPF220, SLy9 and SFHo, which have an intermediate stiffness, receive the highest relative posterior densities, whereas NL3, BMPF275 and BMPF260, some of the stiffest EoSs,  receive the lowest. Since many of the corresponding $M(R)$ relations overlap with the current observational uncertainties, these rankings should not be interpreted as a definitive preference for specific nuclear models, but rather as a qualitative demonstration that the learned posterior distinguishes between different candidate EoSs.

Overall, these experiments demonstrate that the proposed framework can naturally accommodate current datasets containing an arbitrary number of observations while remaining sensitive to physically meaningful changes in both the observational uncertainties and the underlying $M(R)$ relations. Quantitative inference using current NICER and the LVK observations, however, will require training on more realistic observational uncertainty models that accurately reproduce the structure of the measured posteriors.

\section{Conclusions}
\label{conclusion}
We have presented NS-UNO, to the best of our knowledge, the first NPE framework designed to infer the NS EoS from an unconstrained number of observations while preserving the posterior samples associated with each measurement. NS-UNO achieves this through a hierarchical DeepSets architecture that reflects the two-level structure of the observational data: a set of NS observations, each itself represented by a set of samples. The architecture is permutation-invariant at both levels and maps observational sets of varying size into a fixed dimensional context representation that conditions a CNF, allowing a single trained model to infer the posterior distribution of the EoS for different observational configurations.

The model was trained jointly on PT and GP EoS datasets, exposing the network to substantially different descriptions of dense matter within a single inference framework. Across both test sets, the model provides accurate EoS reconstructions together with well-calibrated posterior uncertainties. The reconstruction is most accurate within the density range directly probed by the available NS observations, while the uncertainty and reconstruction error increase beyond the maximum central density reached by the observed stars. This behaviour reflects the decreasing amount of observational information available to constrain the EoS at higher densities.

By systematically varying the observational configuration, we further showed that the mass range covered by the observations is more important than simply increasing their number. Although additional observations generally improve the reconstruction, the gain becomes progressively smaller as $N_{\rm obs}$ increases. In contrast, observations distributed over a broad range of NS masses provide complementary information about different density regimes of the EoS. In particular, the inclusion of high-mass NSs extends the observational sensitivity towards higher densities and improves the reconstruction of the high density EoS. These results emphasize that the constraining power of future datasets will depend not only on the number and precision of detected sources, but also on the region of the $M(R)$ relation that they probe.

We additionally assessed the ability of the model to generalize beyond the EoS ensembles used during training by considering several microscopic and phenomenological EoS models. The framework successfully reconstructs a broad range of previously unseen EoSs, although the reconstruction quality deteriorates for models with more extreme $M(R)$ relations. Notably, the hyperonic EoSs DD2Y and SFHoY are reconstructed with an accuracy comparable to their nucleonic counterparts, despite the training data not explicitly encoding microscopic particle composition. This demonstrates that the framework can generalize to physically motivated EoSs with properties and compositions not explicitly represented as distinct model classes during training.

Finally, we investigated the behaviour of the learned posterior outside the observational conditions represented during training. The model responds systematically to increasing observational uncertainties and can directly process posterior samples from current multimessenger measurements. When conditioned on current NICER and the LVK GW170817 observations, it assigns different posterior densities to candidate theoretical EoSs in a manner qualitatively consistent with their $M(R)$ relations and the observational constraints. This application should not be interpreted as a quantitative EoS inference or model selection analysis, but demonstrates that the architecture can operate directly on realistic posterior samples and remains sensitive to physically meaningful differences between candidate EoSs.

The present work is primarily a methodological proof of concept and several limitations remain. In particular, the training observations are represented by correlated Gaussian distributions with uncertainty ranges that are generally narrower and structurally simpler than the posteriors of current NICER and LVK measurements. Quantitative application to present observations will therefore require training on more realistic uncertainty models, including non-Gaussian posterior structures, which will be exploited in future work. Furthermore, the present implementation assumes the same set of observables for each observation. Although the hierarchical set based architecture provides a natural basis for combining heterogeneous measurements, this capability remains to be explicitly developed and tested.

Several extensions are therefore particularly relevant. A natural next step is the inclusion of additional observables, most notably tidal deformabilities from GW measurements, enabling a fully multimessenger implementation in which different observations may provide different combinations of astrophysical information. The treatment of observational uncertainties should also be extended to encompass the broader range and complexity expected from both current and future measurements. Finally, EoSs containing phase transitions remain challenging for the present model to reconstruct reliably.

Overall, NS-UNO provides a flexible and scalable approach to NS EoS inference that is not tied to a fixed number of observations or to a single EoS parametrization. By operating directly on sets of posterior samples and accommodating observational datasets of varying size, it provides a natural foundation for exploiting the increasingly numerous and heterogeneous multimessenger observations expected from the next generation of NS experiments.

\section*{Acknowledgements} 

V.C. expresses sincere gratitude to the FCT for their generous support through Ph.D. grant number 2024.00311.BD. This work was partially supported by national funds from FCT (Fundação para a Ciência e a Tecnologia, I.P, Portugal) under the projects 2022.06460.PTDC with the  DOI identifier 10.54499/2022.06460.PTDC, and UIDB/04564/2020 and UIDP/04564/2020, with DOI identifiers 10.54499/UIDB/04564/2020 and 10.54499/UIDP/04564/2020, respectively, by the European Union-Next Generation EU, Mission 4 Component 1 CUP J53D23001550006 with the PRIN Project No. 202275HT58, and the Polish National Science Center OPUS grant no. 2021/43/B/ST9/01714. Computing resources were partially provided by the Nicolaus Copernicus Astronomical Center of the Polish Academy of Sciences.

\newpage

\onecolumngrid
\appendix

\section{Sensitivity to phase transition like signatures} \label{app:phaseTransition}

\begin{figure}[!hbt]
    \centering
    \includegraphics[width=0.47\linewidth]{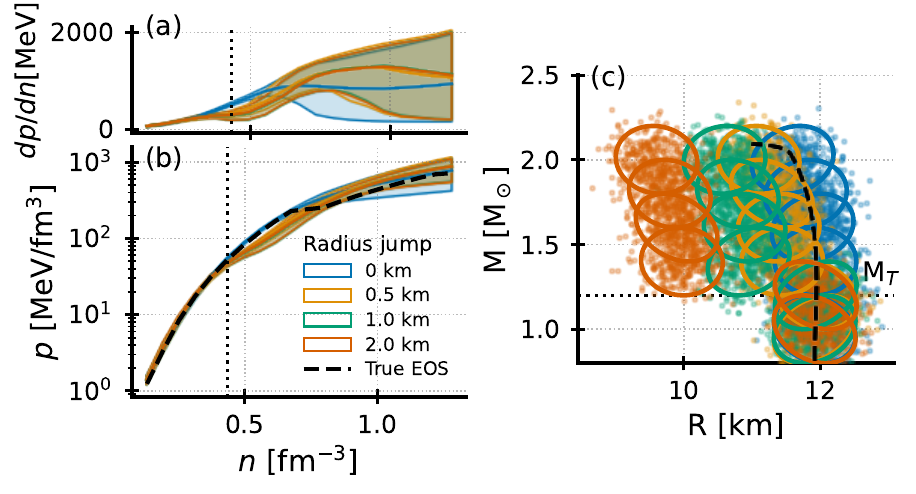}
    \hskip 0.05\linewidth 
    \includegraphics[width=0.47\linewidth]{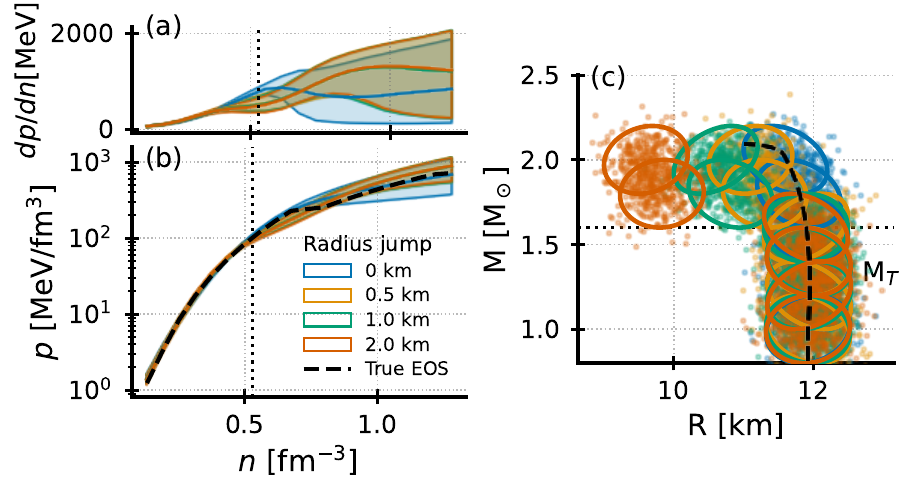}
\caption{
Sensitivity of the inferred EoS to artificial phase transition like signatures introduced in the $M(R)$ relation. The left and right panels correspond to transition masses $M_T=1.2\,M_\odot$ and $M_T=1.6\,M_\odot$, respectively. Above $M_T$, the stellar radius is artificially reduced by 0, 0.5, 1.0, and 2.0 km (blue, yellow, green, and orange), and each inference is performed using six mock observations. Panels (a) show the inferred pressure gradient, $dp/dn$, panels (b) the reconstructed EoS together with the true EoS (black dashed curve), and panels (c) the corresponding $M(R)$ observations. The vertical dashed lines in panels (a) and the horizontal dashed lines in panels (c) indicate the transition density and transition mass, respectively.
}
    \label{fig:Phase_trans1}
\end{figure}

To test the limits of generalization capabilities of the model, we investigate whether the model is sensitive to modifications of the $M(R)$ relation resembling signatures that may arise from a first-order phase transition. Following the approach adopted in \cite{carvalho2024neutron}, we introduce an artificial discontinuous reduction in the stellar radius above a chosen transition mass and analyse its impact on the inferred EoS. Figure~\ref{fig:Phase_trans1} presents the results for transition masses of $1.2\,M_\odot$ (left plots) and $1.6\,M_\odot$ (right plots). Radius jumps of 0, 0.5, 1.0, and 2.0 km are shown in blue, yellow, green, and orange, respectively. Each reconstruction is performed using six mock observations. The upper plot shows the inferred pressure slope, $dp/dn$, the lower plot the reconstructed pressure as a function of baryon density, and the right plot the corresponding $M(R)$ observations.

The results show that the inferred EoS responds systematically to the artificial modifications of the $M(R)$ relation. As the imposed radius jump increases, corresponding changes appear in the reconstructed EoS and its pressure gradient. In particular, $dp/dn$ exhibits a localized reduction around the central density associated with the imposed transition mass, indicated by the vertical dashed line. This feature shifts towards higher densities when the transition mass is increased from $1.2\,M_\odot$ to $1.6\,M_\odot$. This experiment therefore suggests that the model is sensitive to localized modifications of the $M(R)$ relation and can reflect them in the corresponding density region of the inferred EoS.

We additionally tested the framework on EoSs containing explicit phase transitions. In these cases, the model was not able to reliably reconstruct the sharp changes in the EoS associated with the transition. Therefore, the results presented here should only be interpreted as a test of the model sensitivity to phase transition like modifications of the $M(R)$ relation, rather than as evidence that the framework can accurately reconstruct EoSs containing first order phase transitions. Extending the training data to better represent such features and investigating their reconstruction will be addressed in future work.

\section{Implementation Details}
\label{ssec:implementation}

Our architecture uses the following components:
\begin{itemize}
\item \textbf{Software}: Implemented in \texttt{PyTorch} \cite{NEURIPS2019_9015} with the \texttt{nflows} library \cite{nflows}.
\item \textbf{Flow Design}:
\begin{itemize}
\item Number of flow transformations: 16,
\item 3 ResNet blocks \cite{he2016deep} per flow with 120 hidden units,
\item Exponential Linear Unit (ELU) activation  functions \cite{clevert2015fast} for smooth gradients.
\end{itemize}
\item \textbf{Optimization}:
\begin{itemize}
\item Adam optimizer \cite{kingma2014adam} with an initial learning rate of $10^{-3}$ and weight decay $10^{-5}$. The learning rate is reduced by a factor of 10 every 800 epochs using a StepLR scheduler.
\item Batch size of 384 and training for 2000 epochs.
\item We set the strength of the physical regularization $\lambda = 0.7$ (see Eq.~\ref{eq:total_loss} for the definition of the total training loss) based on empirical optimisation.
\end{itemize}
\end{itemize}

\twocolumngrid

\bibliographystyle{apsrev4-1}
\bibliography{biblio}

\end{document}